\documentclass[journal]{IEEEtran}

\usepackage{cite}
\ifCLASSINFOpdf
   \usepackage[pdftex]{graphicx}
		\usepackage{epstopdf}
\else
   \usepackage[dvips]{graphicx}
\fi
\usepackage{amsmath}
\ifCLASSOPTIONcompsoc
  \usepackage[caption=false,font=normalsize,labelfont=sf,textfont=sf]{subfig}
\else
  \usepackage[caption=false,font=footnotesize]{subfig}
\fi

\usepackage{siunitx}
\usepackage{tikz}
\usetikzlibrary{patterns}

\begin{document}
\title{Analytical Drain Current Model of One-Dimensional Ballistic Schottky-Barrier Transistors}

\author{Igor~Bejenari,
       Michael Schr\"oter, ~\IEEEmembership{Senior Member,~IEEE,}
        and~Martin~Claus % <-this % stops a space
				\thanks{Manuscript received October xx, 2016; revised December xx, xx. This work was supported in part by a grant from the Cfaed, CAPES project 88881.030371/2013-01, DFG project CL384/2, and  DFG project SCHR695/6.}%
\thanks{I. Bejenari, M. Schr\"oter  and M. Claus are with the Chair for Electron Devices and Integrated Circuits, Department of Electrical and Computer Engineering, 
 Technische Universit\"at Dresden, 01062, Germany.}%
\thanks{I. Bejenari is  also with Institute of Electronic Engineering and Nanotechnologies,  Academy of Sciences of Moldova, MD 2028 Chisinau, Moldova (e-mail:igor.bejenari@fulbrightmail.org).}% <-this % stops a space
\thanks{M. Schr\"oter is also with the Department of Electronics
and Communication Engineering, University of California at San Diego,
La Jolla, CA 92093 USA (e-mail:Michael.Schroeter@tu-dresden.de).}
% <-this % stops a space
\thanks{M. Claus is also  with the Center for Advancing Electronics Dresden (Cfaed),
 Technische Universit\"at Dresden, 01062, Germany (e-mail:Martin.Claus@tu-dresden.de).}
% <-this % stops a space
}%


\markboth{IEEE TRANSACTIONS ON ELECTRON DEVICES,~Vol.~x, No.~xx, December~xx}%
{Bejenari \MakeLowercase{\textit{et al.}}: Analytical Model of One-Dimensional Ballistic Schottky-Barrier Transistors}

%\IEEEpubid{ 0000--0000/00\$00.00~\copyright~2015 IEEE}
%\IEEEpubid{"©2017 IEEE. Personal use of this material is permitted. Permission from IEEE must be obtained for all other uses, including reprinting/republishing this material for advertising or promotional purposes, collecting new collected works for resale or redistribution to servers or lists, or reuse of any copyrighted component of this work in other works."}
%\IEEEpubid{\makebox[\columnwidth]{\textcopyright 2017 IEEE. Personal use of this material is permitted. Permission from IEEE must be obtained for all other uses, \hfill} \hspace{\columnsep}\makebox[\columnwidth]{ }}
\IEEEpubid{\begin{minipage}{\textwidth}\ \\[12pt] \hfill
\copyright 2017 IEEE. Personal use of this material is permitted. \\ Permission from IEEE must be obtained for all other uses, including reprinting/republishing this material for advertising or promotional purposes, collecting new collected works for resale or redistribution to servers or lists, or reuse of any copyrighted component of this work in other works.
\end{minipage}}



\maketitle
\begin{abstract}
A new  analytical model based on the WKB approximation for MOSFET-like one-dimensional ballistic transistors with Schottky-Barrier  contacts has been developed  for the drain current. By using a proper approximation of both the Fermi-Dirac distribution function and transmission probability, an analytical solution for the Landauer integral was obtained, which overcomes the limitations of existing models and extends their applicability toward high bias voltages needed for analog applications.
%The model has been justified by a self-consistent Schrodinger-Poisson equations solver.
The simulations of transfer and output characteristics are found to be in agreement with the experimental data for sub 10 nm CNTFETs.
\end{abstract}


% \fi
\IEEEpeerreviewmaketitle

\section{Introduction}
\IEEEPARstart{S}{ome} of the recent requirements for CMOS technology listed in the International
Roadmap for Devices and Systems (IRDS) \cite{IRDS} include  high--mobility channel materials, 
gate--all--around (nanowire) structures, scaling down  supply voltages lower than 0.6 V, controlling source/drain series resistance within tolerable limits, and fabrication of advanced nonplanar multi--gate and nanowire MOSFETs with gate lengths below 10 nm. 
Along with FETs based on semiconductor nanowires, carbon-nanotube FETs (CNTFETs) satisfy these requirements \cite{Guo_IEEE2004, Franklin_2012,Peng2017}. 
Downscaling the transistor dimensions goes along with a transformation of ohmic contacts into Schottky contacts~\cite{Larson_IEEE2006,Leonard2011}.
Due to a possible low channel resistance (or even ballistic conduction), the metal-semiconductor contact
resistance can significantly affect or even dominate the performance of Schottky--barrier (SB) transistors~\cite{Chen_IEEE2006,Heinze2003,Chen2005}.

Unlike conventional transistors with ohmic contacts, SB transistors operate not only by varying the
channel potential but also by changing the shape of the SB, which induces a voltage dependence in the contact resistance due to the variation of the source and drain transparency.
The latter depends on the type of metal contact (i.e. Pd, Rh, Pt, Ti or
Al), the surface preparation, annealing  conditions, the CNT diameter, and the electrostatic potential~\cite{Kim2005,Chen2005,Fediai2016}.
Moreover, an external bias modulates the dimensional characteristics (height and width) and thus the  transparency, of the SBs located at the interfaces between the internal channel and S/D metal contacts~\cite{Knoch2008,Svensson2011}.

For circuit design, the description of the device behavior based on the nonequilibrium Green’s function (NEGF) method, Wigner transport equation, and Boltzmann  equation formalism is unsuitable in terms of memory and time~\cite{Guo_IEEE2004,Ossaimee_EL2008,Leonard_Nanotech2006,Maneux_SSE2013}.
To reduce the computation time, simplified semi--analytical models have been proposed to analyze the ${I-V}$ characteristics of quasi-1D SB-FETs solving the current integral involved in the transport calculations numerically \cite{Hazeghi_IEEE2007,Jimenez_Nanotech2007,Michetti_IEEE2010,Sinha_IEEE2009,Zhu_IEEE2009,Vega_IEEE42006,Vega_IEEE72006,Vega_IEEE92006}.
Even though these models are a very good compromise between accuracy and computational efficiency, they are still not suitable for circuit simulation in a SPICE-like environment.
For practical circuit design, compact models are required.
Different simple analytical expressions for the drain current obtained by using an energy independent transmission probability have been reported in the literature for SB-FETs~\cite{Kazmierski_IEEE2010,Wong_IEEE2007PII,Najari_IEEE2011,Maneux_SSE2013,Ragi_IEEE2016}.
In these models, the simulated ${I-V}$ characteristics agree with experimental data in a limited bias range~\cite{Maneux_SSE2013,Zhu_IEEE2010}.
Also, the simulation results do not agree well with those based on a numerical solution of the current integral.
The semi--empirical virtual source model based on a set of empirical fitting parameters is reliable in the framework of experimental data~\cite{Wong_IEEE2013,Wong_IEEE2015PI,Wong_IEEE2015PII},
but it cannot be used for predictions.
%end

%This paper is  intended to describe the ${I-V}$ characteristics of MOSFET-like one-dimensional transistors with Schottky-Barrier  contacts based on the analytical transport model free of numerical computation of the current integral.
In this paper, we demonstrate the analytical drain current model free from numerical computation of the current integral. It allows to simulate  ${I-V}$ characteristics of MOSFET-like one-dimensional transistors with SB  contacts with  reduced computation time.
We adopt the pseudo-bulk approximation~\cite{Fregonese_IEEE2008} to self--consistently estimate the channel potential variation under applied bias with respect to channel charge.
The  drain--current model captures a number of features such as ballistic transport, transmission through the SB contacts,  and ambipolar conduction.
It can be applied to quasi-1D SB-FETs based on both nanowires and nanotubes at large bias voltages. 
The proposed analytical model can replace the transport equations in empirical-based compact models~\cite{Schroter_IEEE2015,nanohub2015} to improve their applicability.

 
%hfill August 26, 2015

\section{Transport Model}
\IEEEpubidadjcol
 \subsection {Energy Band Model} 
The band model used was adopted from the evanescent mode analysis approach~\cite{Oh_IEEE2000,Michetti_IEEE2010,Jimenez_Nanotech2007,Hazeghi_IEEE2007}.
The electrostatic potential, ${\psi(r)}$, inside a transistor contains 
a transverse potential ${\psi_{t}(r)}$, which describes the electrostatics perpendicular to the channel and represents a partial solution of Poisson's equation,
as well as a longitudinal potential ${\psi_{l}(r)}$ called evanescent
mode, responsible for the potential variation along the channel. 
The transverse potential inside the channel is reduced to ${\psi_{t}(r)\approx \psi_{cc}}$, where ${\psi_{cc}}$ is the channel (surface) potential at the current control point~\cite{Lundstrom_IEEE2003,Mothes2015}.
The longitudinal solution ${\psi_{l}(r)}$ is obtained solving the Laplace equation along the transport direction.
Therefore, near the source and drain contacts, the conduction subband edge in the left ($E^{\rm{c}}_{\rm{L}}$) and right ($E^{\rm{c}}_{\rm{R}}$) regions can be  given by 
\begin{eqnarray}
E^{\rm{c}}_{\rm{L}}(z)=E_{m,0}-q\psi_{cc}  + E^s_b\exp{\left(-\frac{z}{\lambda}\right)},
\label{eq:SBHight_1} \\
E^{\rm{c}}_{\rm{R}}(z)=E_{m,0}-q\psi_{cc}  + E^d_b \exp{\left(\frac{z-L}{\lambda}\right)},
\label{eq:SBHight_2} 
\end{eqnarray}
where $L$ is the total length of the channel, $\lambda$ is a characteristic length of the decaying electrostatic potential that can be interpreted as an effective SB width and ${E^{s(d)}_b=\phi_{b}+q\psi_{cc}-E_{m,0}-qV_{s(d)}}$ is the bias dependent potential barrier height with respect to the bottom of the $m$th conduction subband ${E_{m,0}-q\psi_{cc}}$ at the source and drain contacts, correspondingly.
For cylindrical gate-all-around FETs, the asymptotic value of $\lambda$ is approximately given by ${(2\kappa t_{\rm{ox}}+d_{\rm{ch}})/4.81}$, where ${\kappa=\epsilon_{\rm{ch}}/\epsilon_{\rm{ox}}}$ can be obtained if the oxide thickness, $t_{\rm{ox}}$, is significantly smaller than the channel diameter, ${d_{\rm{ch}}}$~\cite{Oh_IEEE2000}.
For gate-all-around CNTFETs, the CNT diameter, ${d_{\rm{CNT}}}$, is often smaller than the oxide thickness, therefore, the asymptotic value of ${\lambda}$ is slightly modified~\cite{Wong_IEEE2015PI}. In the case of double-gate FETs, the similar approximation of the characteristic length reads ${\lambda\approx (2\kappa t_{\rm{ox}}+t_{\rm{ch}})/\pi}$, where  ${t_{\rm{ch}}}$ is the thickness of the channel~\cite{Oh_IEEE2000}.

\begin{figure}[!t]
\centering
\includegraphics[width=3.5in]{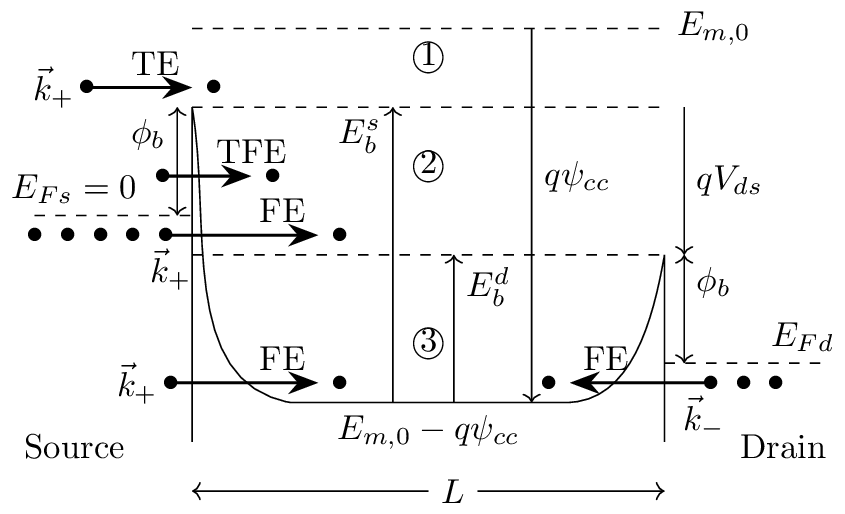}
\caption{Energy band diagram describing  the thermionic emission (TE) current in region 1, thermionic--field emission (TFE) and field emission (FE)  tunneling from source into region 2, as well as FE tunneling from both source and drain into region 3. The TE and TFE tunneling from drain into channel are not shown. At equilibrium, the Fermi level ${E_{Fs}}$  of the source contact is in the middle of the tube bandgap ${E_g}$.}
\label{fig1:BandDiagram}
\end{figure}

Fig.~\ref{fig1:BandDiagram} shows the conduction band profile along the channel.
The gate length ${L_g}$ of the device coincides with the channel length ${L}$. 
The metal-semiconductor barrier height referenced to source Fermi level $E_{Fs}$ is described by a bias independent parameter, ${\phi_{b}}$, which is commonly defined by the difference between the metal work function, $\phi_M$, and semiconductor electron affinity, $\chi_{SC}$, i.e., $\phi_{b}\approx\phi_M-\chi_{SC}$~\cite{Sze_2007,Svensson2011,Tung2014}.
For holes, the similar parameter ${\phi^{h}_{b}}$ is given by ${\phi^{h}_{b}=E_g-\phi_{b}}$, where ${E_g=2E_{m,0}}$ is the band gap.  
The source and drain Fermi levels $E_{Fs}$ and $E_{Fd}$, respectively, are related as $E_{Fd}=E_{Fs}-qV_{ds}$, where $V_{ds}=V_{d}-V_{s}$ is the drain--source voltage.
For the sake of simplicity, we further assume that the source voltage is equal to zero, ${V_s=0}$.
Fig.~\ref{fig1:BandDiagram} describes different injection mechanisms in the device.
Region 1 corresponds to a thermionic current due to thermally excited electrons mainly injected from the source into the channel without reflection. These electrons have enough energy to  overcome the potential barrier. 
The  electrons with energy belonging to region 2 tunnel from the source through the barrier into the channel and propagate towards the drain contact. 
Due to their large energy, these electrons overcome the barrier in the drain region and are absorbed by the drain contact without reflection.
Belonging to the same interval of energy, the thermally excited electrons injected from the drain freely propagate in the channel and can tunnel through the potential barrier located at the source. 
Region 3 is limited from below by the electron subband edge and from above by the top of the drain barrier. 
In this case, the electrons tunnel  from both the source and drain through the SBs into the channel. After multiple reflections between the barriers, these electrons are absorbed in the source or drain contacts. 
The multiple reflection represents a second order contribution to the total current.
%It may be important at large bias.
 
The contribution of electrons injected from the source and drain to the total current depends on both the energy dependent transmission through the channel and electron distribution in the contacts.

\subsection{Piece-Wise Approximation of Fermi-Dirac Distribution Function}

The electron distribution in the source/drain contacts is given by the equilibrium Fermi-Dirac distribution function ${f_{FD}(E-E_F)=1/\left\{\exp\left[\left(E-E_F\right)/{k_BT}\right]+1\right\}}$.
To find an analytical expression for the current, we use a piece-wise approximation for $f_{FD}(E)$ given by
\begin{equation} 
{  f_{\rm{app}}(E)= \left\{
				\begin{array}{l}
					1-\frac{1}{2}\exp\left( \frac{E-E_F}{{c_1 k_BT}}\right),  E \leq E_F  \\
					\frac{1}{2}\exp\left(\frac{E_F-E}{{c_1 k_BT}}\right),~  E_F<E<E_F+c_2 k_BT   \\
				\exp\left(\frac{E_F-E}{{k_BT}}\right), ~   E\geq E_F+c_2 k_BT	
				\end{array}
				\right.\
	\label{eq:Fermi_Dirac}		
}
\end{equation}
where factor ${c_1=2\ln(2)}$ was introduced  to get a similar curvature of $f_{\rm{app}}(E)$ compared to $f_{FD}(E)$. 
The value of ${c_1}$ is defined by requesting the areas under the $f_{\rm{app}}(E)$ and $f_{FD}(E)$ curves to be equal, i.e., ${\int^{\infty}_{E_F} f_{FD}(E-E_F)dE=0.5\int^{\infty}_{E_F} \exp\left[ (E_F-E)/c_1 k_B T \right] dE}$.   
Parameter ${c_2=2\ln^2(2)/(2\ln(2)-1)\approx 2.49}$ is obtained from the requirement for continuity  of $f_{\rm{app}}(E)$ at $E_F+c_2 k_BT$.

The approximation $f_{\rm{app}}(E)$ provides accurate values of the electron distribution function at different temperatures in the whole energy range  with a maximum relative error of about 6-9 percent in the vicinity of Fermi level $E_F$. 

\subsection{Barrier Transmission Probability}

In order to estimate the transparency of the source/drain contacts, we use the transmission probability across each SB obtained in the framework of the Wentzel--Kramers--Brillouin (WKB) approximation.
Using the effective mass (parabolic band) approach, the probability $T^{s(d)}_{WKB}(E)$ for electrons to tunnel through an exponential decaying potential barrier of the kind ${E^s_b\exp{\left(-z/\lambda\right)}}$ or
 ${E^d_b\exp{\left[(z-L)/\lambda\right]}}$ is given by the following expression~\cite{Michetti_IEEE2010}
\begin{eqnarray}
&T^{s(d)}_{WKB}(E)  =\exp\left\{ -\alpha \sqrt{E^{s(d)}_{b}} \gamma \left(E/E^{s(d)}_{b}\right) \right\}, 
\label{eq:Transmission}\\
&\gamma(x)  =\sqrt{1-x}-\sqrt{x}\arctan\left(\sqrt{\frac{1-x}{x}}\right),
\end{eqnarray}
where ${\alpha=4\lambda\sqrt{m^*}/\hbar}$. For CNTs, the electron effective mass is ${m^*=8E_{m,0} \hbar^2/3a^2V_{\pi}^2}$ with $a=\SI{2.49}{\angstrom}$ - carbon--carbon atom distance and $V_{\pi}=\SI{3.033}{\electronvolt}$ -- carbon $\pi-\pi$ bond energy in the tight binding model~\cite{Mintmire1998}.

To obtain an analytical expression for the current, we use the following approximation for $\gamma(x)$ in~(\ref{eq:Transmission})
\begin{eqnarray}
&\gamma_{\rm{app}}(x)  =px-(p+1)\sqrt{x}+1, 
\label{eq:gama_app}\\
&p=(\varphi \gamma(1/\varphi)-\varphi+\sqrt{\varphi})/(1-\sqrt{\varphi})\approx 0.7113,
\end{eqnarray}
where the quantity $\varphi=(1+\sqrt{5})/2$ represents the golden ratio and $x=E/E^{s(d)}_{b}$ is a dimensionless variable.
The absolute error of $\gamma_{\rm{app}}(x)$ is less than 0.02 for all ${x\in[0,1]}$.
Nevertheless, the implementation of $\gamma_{\rm{app}}(x)$ in~(\ref{eq:Transmission}) leads to an increase of relative error of the approximate transmission probability $T^{s(d)}_{\rm{app}}(E)$ with gate voltage due to term ${E^{s(d)}_{b}}$. 
To reduce the relative error, we introduce a correction factor ${\exp\left[\alpha \Delta \sqrt{E^{s(d)}_{b}}\right]}$ with the constant $\Delta < \text{max} \left|\gamma(x)-\gamma_{\rm{app}}(x)\right|$ in the final expression of current.
The approximate transmission probability $T^{s}_{\rm{app}}(E)$ based on~(\ref{eq:Transmission}) and (\ref{eq:gama_app}) is used  in region 2 if there is only one potential barrier. 

The probability of electron transmission through the potential barrier increases with energy. 
At a large gate voltage, electrons with high energy or close to the Fermi level tunnel through the thin potential barrier with a rather small reflection probability ${1-T^{s(d)}_{WKB}(E)}$ and mainly contribute to the current, whereas the contribution of electrons with low energy is not essential due to a small transmission probability ${T^{s(d)}_{WKB}(E)}$. 
Hence, in region 3 (see Fig.~\ref{fig1:BandDiagram}), the multiple reflections between two potential barriers can be neglected.
In this case, the total transmission probability reads
\begin{equation}
T_{2b}(E)=T^{s}_{WKB}(E)T^{d}_{WKB}(E).
\label{eq:total_Transmission}
\end{equation}
The approximate total transmission probability ${T^{2b}_{\rm{app}}(E)}$ can be obtained by using (\ref{eq:Transmission})--(\ref{eq:total_Transmission}).

The presented approach is valid if electron-phonon scattering is relatively small, i.e., the channel length $L$ is of the order of an electron mean free path $L_{\rm{mfp}}$, such that ${L/L_{\rm{mfp}}<1/\overline{T}_b}$, where ${\overline{T}_b}$ is an average value of the SB transmission probability characterizing a source/drain contact transparency~\cite{Knoch2008}.
Depending on the applied bias, the mean free path $L_{\rm{mfp}}$ can vary from 60 to 200 nm~\cite{Fuller2014,Franklin_2010,Zhang2008_nl,Purewal2007,Yao2000} at room temperature in CNTFETs. 
Also, the model does not take into account direct source-to-drain tunneling and short--channel effects (e.g., SS degradation and Drain-Induced Barrier Lowering), which are determined purely by electrostatics and essentially affect the current at ${L\approx\lambda}$~\cite{Knoch2008}.
To extend our model for an analysis of experimental data for short-channel devices, a semi--empirical drain current formulation for the subthreshold region has been introduced in Section III.   

\subsection{Total Current}

To calculate the total electron current, we use the Landauer-Buttiker  approximation for a one-dimensional system~\cite{Datta_1995}
\begin{equation}
I = \frac{4q}{h}  {\int\limits_{0}^\infty T_{WKB}(E)\left[ f_{FD}(E-E_{Fs}) - f_{FD}(E-E_{Fd}) \right]dE},
\label{eq:current}
\end{equation}
where the product of the spin and electron subband degeneracies gives a factor of 4 in front of the integral (\ref{eq:current}) for CNTFETs. 
Here, the Fermi level in the source (drain) contact is defined as $E_{Fs(d)}=q\psi_{cc}-E_{m,0}-qV_{s(d)}$ with reference to the bottom of the conduction subband in the channel.

Referring to the energy band diagram in Fig.~\ref{fig1:BandDiagram}, we can categorize the components of source current into three types: (\textit{i}) thermionic emission (TE) over the potential barrier in region 1, i.e., $T_{WKB}(E)=1$, (\textit{ii}) field emission (FE) below the source Fermi level $E_{Fs}$  in regions 2 and 3, and (\textit{iii}) thermionic-field emission (TFE) through the potential barrier at the energy between TE and FE in region 2.
The components of drain current also can be divided into these three types: (\textit{i}) TE over the potential barrier in region 1, (\textit{ii}) FE tunneling below the drain Fermi level $E_{Fd}$ in region 3, and (\textit{iii}) TFE tunneling  at the energy between TE and FE in regions 2 and 3.
We neglect the source-to-drain and band-to-band tunneling here.

The source (drain) component of TE current is defined as
\begin{equation}
I^{s(d)}_{TE} = \frac{4q}{h} k_B T \ln{\left[1+\exp\left(-\frac{E^{s(d)}_{TE}}{k_B T}\right)\right]},
\label{eq:current_TE}
\end{equation}
where ${E^{s(d)}_{TE}}$ sets the minimum energy, with reference to the source (drain) Fermi level, required for thermionic emission of electrons from the source (drain) into the channel. 
For a small gate voltage ${q\psi_{cc}\leq E_{m,0}-\phi_{b}}$, it coincides with the bottom of the conduction subband with reference to the source (drain) Fermi level, i.e., ${E^{s(d)}_{TE}=E_{m,0}-q\psi_{cc}+qV_{s(d)}}$. For a larger gate voltage ($E^{s}_{b}>0$),  ${E^{s(d)}_{TE}=\phi_{b}+qV_{s(d)}}$.

To obtain analytical expressions of the TFE and FE current at ${q\psi_{cc} > E_{m,0}-\phi_{b}}$, we use both the approximate electron distribution function $f_{\rm{app}}(E)$ and approximate transmission probability ${T_{\rm{app}}(E)=\exp(-AE+B\sqrt{E}-C)}$ in (\ref{eq:current}).
If ${E^d_b<E<E^s_b}$, electrons tunnel through the SB located at the source and the argument of ${T_{\rm{app}}(E)}$ is defined by ${A=\alpha p/\sqrt{E^s_b}}$, ${B=\alpha (p+1)}$, and ${C=\alpha\sqrt{E^s_b}}$. 
If ${E<E^d_b}$, electrons tunnel through both SBs located at the source and drain and the argument is  determined by ${A=\alpha p(1/\sqrt{E^s_b}+1/\sqrt{E^d_b})}$, ${B=2\alpha (p+1)}$, and ${C=\alpha(\sqrt{E^s_b}+\sqrt{E^d_b})}$. 
The contribution of electrons within the energy range ${[E_1,E_2]}$ to the source (drain) FE current reads 
\begin{multline}
  I^{s(d)}_{FE}(E_1,E_2)= \exp{\left(C \Delta\right)}\left[i_0\left(A,E_1\right)-i_0\left(A,E_2\right) \right.\\
					-\frac{1}{2}\exp{\left( \frac{E_1-E_{Fs(d)}}{{c_1 k_BT}}\right)} i_0\left(A-\frac{1}{{c_1 k_BT}},E_1\right)\\
				\left.	+\frac{1}{2}\exp\left( \frac{E_2-E_{Fs(d)}}{{c_1 k_BT}}\right) i_0\left(A-\frac{1}{{c_1 k_BT}},E_2\right)\right], 
					\label{eq:current_FE}
\end{multline}
%where term ${i_0\left(A,E_{1,2}\right)}$ describes the current for degenerate Fermi gas of electrons with an energy greater than $E_{1,2}$
\begin{equation} 
{  i_{0}(y,E)= \left\{
				\begin{array}{l}
					\frac{4qT_{\rm{app}}(E)}{ha}\left[ \frac{B}{2} \sqrt{\frac{\pi}{y}}r\left( \sqrt{y E}-\frac{B}{2\sqrt{y}}\right)+1\right],  y > 0  \\
					\frac{8q}{h B^2}T_{\rm{app}}(E)\left[1 - B \sqrt{E}\right],~ y = 0   \\
				\frac{4qT_{\rm{app}}(E)}{h\left|y\right|}\left[ \frac{B}{\sqrt{\left|y\right|}}F\left( \sqrt{\left|y\right| E}+\frac{B}{2\sqrt{\left|y\right|}}\right)-1\right],  y < 0
				\end{array}
				\right.\
	\label{eq:Current_deg}		
}
\end{equation}
where $y$ is a variable and  ${r(x)=\exp(x^2)\rm{erfc}(x)}$, such that ${r(-x)=2\exp(x^2)-r(x)}$. Both Dawson's integral ${F(x)=\exp{(-x^2)}\int^x_0{\exp{(t^2)}dt}}$ and the function ${r(x)}$ can be calculated by using elementary approximations~\cite{Lether1990,Lether1991,StansLib2012}.
The source (drain) TFE current due to electrons within the energy range ${[E_1,E_2]}$ is given by
\begin{multline}
  I^{s(d)}_{TFE}(E_1,E_2)= 
					\left[e^{\left( \frac{E_{Fs(d)}-E_1}{{c_1 k_BT}}\right)} i_0\left(A+\frac{1}{{c_1 k_BT}},E_1\right)\right.\\
					\left.-e^{\left( \frac{E_{Fs(d)}-E_2}{{c_1 k_BT}}\right)} i_0\left(A+\frac{1}{{c_1 k_BT}},E_2\right)\right] \frac{e^{C \Delta}}{2}. 
					\label{eq:current_TFE}
\end{multline}
If ${E_1>E_{Fs(d)}+c_2k_BT}$, then both the factors $c_1$ in front of thermal energy ${k_BT}$ and $1/2$ in front of correction term ${\exp(C\Delta)}$ are replaced by 1 in (\ref{eq:current_TFE}).

At a small gate voltage or large drain-source voltage ${V_{ds}}$, i.e., ${q\psi_{cc} \leq E_m(0)-\phi_{b}+qV_{ds}}$, there is only SB located at the source (${E^s_b > 0}$ and ${E^d_b\leq 0}$). In this case, the source and drain components of the total current are 
%\begin{equation} 
%\begin{eqnarray}
%&I^s=\left\{
				%\begin{array}{l}
				%I^{s}_{FE}(0,E_{Fs})+I^{s}_{TFE}(E_{Fs},E^s_b)+I^{s}_{TE},~  E_{Fs} \geq 0 \\
					%I^{s}_{TFE}(0,E^s_b)+I^{s}_{TE}, ~ ~ ~E_{Fs} < 0  
			%\end{array}
					%\label{eq:current_s}	
		%\right.\ 	\\
%&I^{d}=I^{d}_{TFE}(0,E^s_b)+I^{d}_{TE}.	
	%\label{eq:current_d}		
%%\end{equation}
%\end{eqnarray}
%\begin{equation} 
\begin{equation}
I^s=\left\{
				\begin{array}{l}
				I^{s}_{FE}(0,E_{Fs})+I^{s}_{TFE}(E_{Fs},E^s_b)+I^{s}_{TE},~  E_{Fs} \geq 0 \\
					I^{s}_{TFE}(0,E^s_b)+I^{s}_{TE}, ~ ~ ~ ~ ~ ~E_{Fs} < 0  
			\end{array}		
		\right.\ 	
		\label{eq:current_s}	
\end{equation}	
\begin{equation}	
I^{d}=I^{d}_{TFE}(0,E^s_b)+I^{d}_{TE}.	
	\label{eq:current_d}		
\end{equation}
Two SBs affect the current for ${q\psi_{cc} > E_m(0)-\phi_{b}+qV_{ds}}$, i.e., ${0<E^d_b<E^s_b}$. In this case, the drain and source components of the total current are given by
\begin{multline}
 I^s =  \theta(-E_{Fs})\left[I^{s}_{TFE}(0,E^d_b)+I^{s}_{TFE}(E^d_b,E^s_b)+I^{s}_{TE} \right] \\
        +\theta(E_{Fs})\theta(E^d_b-E_{Fs})\left[I^{s}_{FE}(0,E_{Fs})+I^{s}_{TFE}(E_{Fs},E^d_b) \right.\\
				\left. +I^{s}_{TFE}(E^d_b,E^s_b)+I^{s}_{TE}\right] \\
				+\theta(E_{Fs}-E^d_b)\left[I^{s}_{FE}(0,E^d_b)+I^{s}_{FE}(E^d_b,E_{Fs}) \right.\\
				\left. +I^{s}_{TFE}(E_{Fs},E^s_b)+I^{s}_{TE}\right],
 \label{eq:Current_s_2b}	
\end{multline}
\begin{multline}
 I^d =  \theta(-E_{Fd})\left[I^{d}_{TFE}(0,E^d_b)+I^{d}_{TFE}(E^d_b,E^s_b)+I^{d}_{TE} \right] \\
        +\theta(E_{Fd})\left[I^{d}_{FE}(0,E_{Fd})+I^{d}_{TFE}(E_{Fd},E^d_b)  \right.\\
				\left. +I^{d}_{TFE}(E^d_b,E^s_b)+I^{d}_{TE}\right],
 \label{eq:Current_d_2b}	
\end{multline}
%\begin{equation} 
%I^d=\left\{
				%\begin{array}{l}
								%I^{d}_{TFE}(0,E^d_b)+I^{d}_{TFE}(E^d_b,E^s_b)+I^{d}_{TE}, ~ F_{Fd} < 0 \\
				%\left[I^{d}_{FE}(0,E_{Fd})+I^{d}_{TFE}(E_{Fd},E^d_b) \right.\\
				%\left. +I^{d}_{TFE}(E^d_b,E^s_b)+I^{d}_{TE}\right],~ ~ ~ ~~~ F_{Fd} \geq 0 
			%\end{array}
		%\right.\ 	\\
	%\label{eq:Current_s_2b}		
%\end{equation} 
%\begin{equation} 
%I^s=\left\{
				%\begin{array}{l}
				%I^{s}_{FE}(0,E_{Fs})+I^{s}_{TFE}(E_{Fs},E^d_b)+ \\
				%I^{s}_{TFE}(E^d_b,E^s_b)+I^{s}_{TE},~  0<F_{Fs} < E^d_b \\
					%I^{s}_{TFE}(0,E^s_b)+I^{s}_{TE}, ~ F_{Fs} < 0  
			%\end{array}
		%\right.\ 	\\
	%\label{eq:Current_d_2b}		
%\end{equation} 
where ${\theta(x)}$ is Heaviside step function. The total current is
\begin{equation}
I(V_{gs},V_{ds}) = I^s(V_{gs},V_{ds}) - I^d(V_{gs},V_{ds}).
\label{eq:tot_current}
\end{equation}
 Due to the electron--hole symmetry of the band structure in a CNT, the total current ${I_a}$ for ambipolar CNTFETs can be defined as
\begin{equation}
I_a(V_{gs},V_{ds}) = I(V_{gs},V_{ds}) + I(V_{ds}-V_{gs},V_{ds}),
\label{eq:total_current}
\end{equation}
where $V_{gs}$ is a gate voltage.
The first (second) term corresponds to a contribution of electrons (holes) to the total current in (\ref{eq:total_current}).

The analytical expressions (\ref{eq:current_s})--(\ref{eq:total_current}) are  valid in the all bias regions. These are smoothly connected without the need of smoothing functions in the different bias regions.

\section{Subthreshold current for short channel devices}

%The subthreshold slope ${S=\left( {d \log_{10}I}/{d V_g} \right)^{-1}}$, where $V_g$ is a gate voltage, can be interpolated as ${S=\left[\alpha t_{ox} + (k_B T \ln 10)^2\right]^{1/2}}$ with $\alpha$ a fitting parameter \cite{Heinze2003}.

For short-channel devices at a small gate voltage ${V_{gs}}$ in the subthreshold region, the thermionic current approximation is not adequate because of a direct source-to-drain tunneling \cite{Wong_IEEE2013,Wong_IEEE2015PII,Balestra_2013}.
For the sake of simplicity, we develop a semi-empirical model in this case. 

 We expand ${\ln(I)}$ given by (\ref{eq:tot_current}) in a Taylor series in the vicinity of the flat band voltage ${qV_{FB}=E_{m,0}-\phi_{b}}$
\begin{equation}
\ln(I) = \ln\left[I(V_{FB})\right] + \frac{\partial\ln(I)}{\partial V_{gs}}(V_{gs}-V_{FB}),
\label{eq:subth_current}
\end{equation}
where ${I(V_{FB})=I^s_{TE}(\phi_b)-I^d_{TE}(\phi_b+V_{ds})}$ is the total TE current defined at ${V_{FB}}$ .
Using (\ref{eq:current_TE}) and (\ref{eq:subth_current}), we obtain the total current in the subthreshold region at ${\psi_{cc}<V_{FB}}$ as
%\begin{equation}
%I_{st} =  I_0 \exp\left[\ln(10)\frac{(V_{g}-V_{FB})}{SS}\right],
%\label{eq:st_current}
%\end{equation}
\begin{eqnarray}
&I_{sub} =  I_0 \exp\left[\ln(10)\frac{(V_{gs}-V_{FB})}{SS}\right], 
\label{eq:st_current}\\
& I_0=\frac{4q}{h} k_B T \ln{\left\{\frac{1+\exp\left(-{\phi_{b}}/{k_B T}\right)}{1+\exp\left[-({\phi_{b}+qV_{ds}})/{k_B T}\right]}\right\}},
\end{eqnarray}
where the subthreshold slope ${SS=\left( {d \log_{10}I}/{d V_{gs}} \right)^{-1}}$ can be approximated by simple expressions including fitting parameters \cite{Heinze2003,Knoch2008,Wong_IEEE2013,Wong_IEEE2015PI} or extracted from either the experimental data \cite{Franklin_2012,Peng2017} or numerical calculations \cite{Leonard_Nanotech2006}. 
The subthreshold current, ${I_{sub}}$, is smoothly connected to the TFE current at ${V_{FB}}$.

\section{Results}

\begin{figure}[!t]
\centering
\includegraphics[width=3.5in]{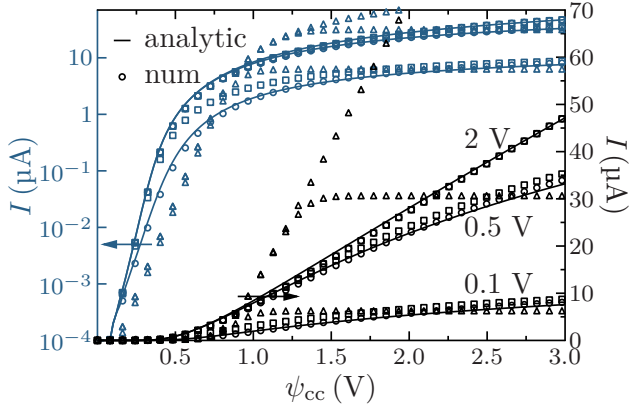}
\caption{Total current $I$ calculated analytically and numerically
as a function of tube potential ${\psi_{cc}}$ at the drain--source voltage ${V_{ds}}$ equal to
0.1, 0.5, and 2 V. CNT chirality (17, 0), bandgap ${E_g = 0.647}$ eV, CNT diameter ${d_{\rm{CNT}}=1.33}$ nm, SB hight ${\phi_b}=E_g/2$, characteristic length ${\lambda=3}$ nm, gate length ${L_g = 20}$ nm, correction parameter ${\Delta=0.0045}$ %${\Delta=4\cdot 10^{-3}}$
 and temperature ${T = 300}$ K. The empty squares correspond to numerically obtained data taking into account  multiple reflections of electron between the SBs~\cite{Hazeghi_IEEE2007}. The empty triangles relate to analytical drain current calculations  within an energy independent effective SB approximation~\cite{Najari_IEEE2011} at the tunneling distance ${d_{\rm{tun}} = 1.5}$ nm and ${V_{ds}=}$0.1, 0.5, and 2 V.}
\label{fig:I_psi}
\end{figure}
We compare the transport characteristics of SB-CNTFETs obtained analytically and numerically.
Also, we compare our results with available experimental data.
The total current is analytically obtained by using (\ref{eq:current_s})--(\ref{eq:Current_d_2b}).
For comparison, we numerically calculate the total current $I$ solving (\ref{eq:current}) with (\ref{eq:Transmission}) and (\ref{eq:total_Transmission}).

Fig.~\ref{fig:I_psi} shows the total current $I$ calculated analytically and numerically  as a function of tube
potential  ${\psi_{cc}}$ at different values of drain--source voltage ${V_{ds}}$
for the SB-CNTFET with a gate length of 20 nm.
For given values of SB height and characteristic length, we have estimated the correction parameter ${\Delta=0.0045}$ using in (\ref{eq:current_FE}) and (\ref{eq:current_TFE}) by a fitting procedure.
Both on linear and logarithmic scale, there is a good agreement between the data obtained analytically and numerically.
The maximum relative error between the analytical and numerical results is about 3 percent at a large bias.
The relative error decreases with a decrease of either the  SB height ${\phi_b}$ or characteristic length ${\lambda}$. 
At a small value of drain--source voltage (${V_{ds}=0.1}$  V), the total current smoothly tends to a constant value with increase of gate bias.
This smooth dependence on the CNT potential is mainly due to the reflection of electrons from both SBs located at the source and drain.
At a larger drain--source voltage (${V_{ds}=0.5}$ and 2 V), the total current strongly depends on the gate voltage in the whole interval of ${\psi_{cc}}$, because the contribution of electrons injected from the drain into the channel is negligibly small  due to the large reflection of such electrons from the SB located at the source.

Fig.~\ref{fig:I_psi} also shows data obtained with two prior models.
The first reference model~\cite{Hazeghi_IEEE2007} is based on the numerical calculation of the Landauer integral  taking into account electron  multiple reflections between SBs located at the source and drain.
The second reference model~\cite{Najari_IEEE2011} is based on the energy independent effective SB approach and provides an analytical expression for the drain current.
In this model, the tunneling distance ${d_{\rm{tun}}}$ represents a fitting parameter.
Depending on the applied bias, the source (drain) energy independent effective SB height ${E^{s(d)}_{b,\rm{eff}}}$  with reference to the conduction band edge is given by equation ${E^{s(d)}_{b,\rm{eff}}=E^{s(d)}_{b} \exp(-d_{\rm{tun}}/\lambda)}$.
If $V_{ds}$ is below pinch-off and $V_{gs}$ is in the vicinity of the threshold voltage,
our model slightly underestimates the drain current compared to the first reference model due to the neglection of electron multiple reflections between SBs located at the source and drain.
In all other cases, the effect of electron multiple reflections is negligible and the agreement between our model and the first reference model is pretty good.
By means of fitting to the first reference (numerical) model, the optimum value of ${d_{\rm{tun}}}$ is estimated to be a half of the characteristic length ${\lambda=3}$ nm at different values of bias.
In this case,  ${E^{s(d)}_{b,\rm{eff}}}$ is less than ${E^{s(d)}_{b}}$ by 40 percent. 
The second reference model roughly agrees with the other two models at small values of $V_{ds}$ but displays large differences at large $V_{ds}$.
The reason is that the analytical expression of the drain current in that model, which corresponds to the TE current with a shifted Fermi level and included energy independent transmission probability of electrons, can be used at small $V_{ds}$.
 The contribution of the TE current to the total current is negligible compared to the contribution of the TFE and FE currents at larger $V_{ds}$, so that it cannot adequately describe the drain current.
The comparison of the output characteristics obtained in the different models gives similar conclusions.
The other prior analytical drain current models~\cite{Kazmierski_IEEE2010,Maneux_SSE2013,Ragi_IEEE2016} have the similar deficiency.
Although these models differently define the energy independent transmission probability  of electrons and Fermi level in the source/drain contacts, they are based on the TE current formulation valid only at small bias. 
Additional simulations based on the numerical integration of the Landauer equation and our analytical drain current equation show  good correspondence for different temperatures, SB heights, and characteristic length. 
Therefore, our model provides results which are in good agreement with the numerical integration of the Landauer integral and is more accurate compared to analytical expressions of the drain current in other existing compact models.
\begin{figure}[!t]
\centering
\includegraphics[width=3.3in]{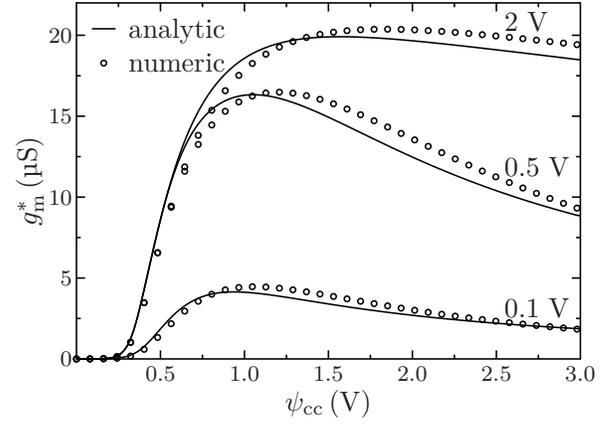}
\caption{Intrinsic transconductance $g^*_m$ calculated analytically and numerically
as a function of tube potential ${\psi_{cc}}$ at the drain--source voltage ${V_{ds}}$ equal to
0.1, 0.5, and 2 V. CNT chirality (17, 0), bandgap ${E_g = 0.647}$ eV, CNT diameter ${d_{\rm{CNT}}=1.33}$ nm, SB hight ${\phi_b}=E_g/2$, characteristic length ${\lambda=3}$ nm, gate length ${L_g = 20}$ nm, correction parameter ${\Delta=0.0045}$ %${\Delta=4\cdot 10^{-3}}$
 and temperature ${T = 300}$ K.}
\label{fig:gm}
\end{figure}

Fig.~\ref{fig:gm} compares the intrinsic transconductance, $g^*_m=\partial I/ \partial \psi_{cc}$, calculated analytically and numerically  at different values of drain--source voltage ${V_{ds}}$
for the SB-CNTFET with a gate length of 20 nm. 
The maximum relative error between the analytical and numerical results is about 8 percent.
The ${g_m}$ reaches its maximum value at ${\psi^{gmax}_{cc}}$, which  mainly depends on the rate of electron reflection from the SBs located at the source and drain.   
\begin{figure}[!t]
\centering
\includegraphics[width=3.3in]{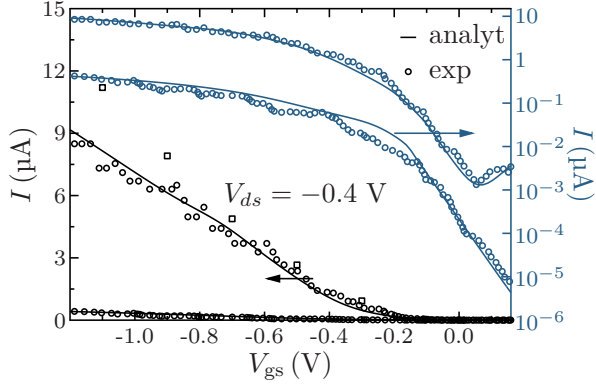}
\caption{Source--drain current $I$ measured (data are from~\cite{Franklin_2012}) and calculated
in the analytical model as a function of gate voltage ${V_{gs}}$ at the drain--source voltage ${V_{ds}}$ equal to
-0.01 and  -0.4 V. The empty squares correspond to measured data extracted from the output characteristics at ${V_{ds}=-0.4}$ V. CNT chirality (17, 0), bandgap ${E_g = 0.647}$ eV, CNT diameter ${d_{\rm{CNT}}=1.33}$ nm, SB hight ${\phi_b}=0.13$ eV, characteristic length ${\lambda=5}$ nm, gate length ${L_g = 9}$ nm, gate oxide capacitance per unit length ${C_{ox}=419}$ aF/um, correction parameter ${\Delta=0.0045}$  and temperature ${T = 300}$ K.}
\label{fig:I_Vg}
\end{figure}
 
In the CNTs with a channel length less than 10 nm, the electron transport is nearly ballistic due to a lack of electron--phonon scattering.
Therefore, to test the validity of our analytical model, we compare the
simulation results with the experimental data for
a bottom gate CNTFET with the gate length ${L_g = 9}$ nm~\cite{Franklin_2012}.
In the subthreshold region of the short--channel device operation, the source--to--drain tunneling and band--to--band tunneling at the drain side cause an increase of the off--current~\cite{Wong_IEEE2015PII}. 
To take into account this effect in our calculations, we replace the TE current~(\ref{eq:current_TE}) by the subthreshold current~(\ref{eq:st_current}) at ${q\psi_{cc}<E_{m}(0)-\phi_{b}}$. 
We use the value of the subthreshold slope $SS=94$ mV/decade extracted from the experimental data in~\cite{Franklin_2012}.
The solution method to self--consistently calculate the CNT potential $\psi_{cc}(V_g)$ was adopted from~\cite{Fregonese_IEEE2008}.
For the bottom gate CNTFETs, the (gate-to-channel) oxide capacitance per unit length is defined as
${C_{ox}=2\pi\epsilon_{0}\epsilon_{r}/ \text{arccosh}(1+2t_{ox}/d_{\rm{CNT}})}$~\cite{Wong_IEEE2007Sep}.
For the ${\rm{HfO_2}}$ gate dielectric with a thickness of ${t_{ox}=3}$ nm and a relative dielectric constant of ${\epsilon_r\approx 18}$, the gate oxide capacitance is equal to 419 aF/um at ${d_{\rm{CNT}}=1.33}$ nm.
For the sake of simplicity, we do not consider the coupling capacitance between the gate and source/drain.

Fig.~\ref{fig:I_Vg} shows the transfer characteristics obtained in the
analytical model compared with the experimental data for the bottom gate p--type CNTFET at the drain--source voltage ${V_{ds}}$ equal to -0.01 and  -0.4 V.
There is a disagreement between the measured transfer characteristic and the data extracted from the output characterisics at ${V_{ds}=-0.4}$ V due to a trap--induced hysteresis in the experimental data~\cite{Haferlach2016}. 
Since the analytical model does not take into account the interface scattering at the non-ideal metal-CNT contact, additional series resistances is needed to fit the experimental data.
In addition, the combination of the quantum mechanical reflection from the potential barrier and electron--phonon scattering  gives a nearly energy independent reduction of the ballistic current across the SB~\cite{Schowalter1991}.
Hence, the current obtained with the analytical model was scaled down by a constant factor of 0.85.
The best agreement between the measurement data and simulation results is found if the bias independent SB height, $\phi_b$, is 0.13 eV and characteristic length, $\lambda$, is 5 nm.
In this case, both $\phi_b$ and $\lambda$ serve as fitting parameters.
At low ${V_g}$, fitting the $I-V$ characteristics can be done by adjusting $\phi_b$, whereas the change of $\lambda$ allows to fit the characteristics at a high bias.  
\begin{figure}[!t]
\centering
\includegraphics[width=3.3in]{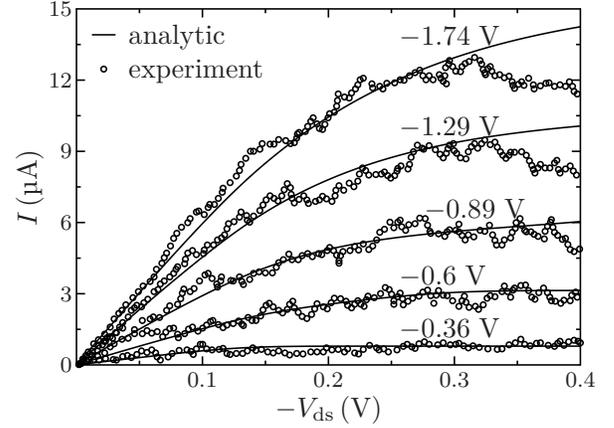}
\caption{Source--drain current $I$ calculated
by the analytical model as a function of drain--source voltage ${V_{ds}}$ at the gate voltage ${V_{gs}}$ equal to
-0.36, -0.6, -0.89, -1.29, and  -1.74 V in comparison with the trap--affected output characteristics (data are from~\cite{Franklin_2012}) measured at ${V_{gs}=}$-0.3, -0.5, -0.7, -0.9, -1.1 V. CNT chirality (17, 0), bandgap ${E_g = 0.647}$ eV, CNT diameter ${d_{\rm{CNT}}=1.33}$ nm, SB hight ${\phi_b}=0.13$ eV, characteristic length ${\lambda=5}$ nm, gate length ${L_g = 9}$ nm, gate oxide capacitance per unit length ${C_{ox}=419}$ aF/um, correction parameter ${\Delta=0.0045}$  and temperature ${T = 300}$ K.}
\label{fig:I_Vds}
\end{figure}

Fig.~\ref{fig:I_Vds} depicts the output characteristics obtained by the
analytical model compared with the experimental data for the bottom gate p--type CNTFET at different gate voltages. 
In comparison with experimental data, the simulated output characteristics are obtained at different values of ${V_g}$ due to the trap--induced hysteresis (see Fig.\ref{fig:I_Vg}).

Our model does not take into consideration the electron band--to--band tunneling and multiple reflections between the SBs located at the source and drain, which lead to a slight increase of the total electron transmission probability to tunnel through the channel at a small bias.
Also, the electron reflections from the SBs are not taken into account in the charge calculations~\cite{Fregonese_IEEE2008} adopted in our model, which would cause a small increase of the tube potential ${\psi_{cc}}$ at a large drain--source voltage.
The validation of the charge model for one-dimensional ballistic Schottky-Barrier transistors represents a separate problem, which requires an additional study.

\section{Conclusion}

An analytical model for ballistic MOSFET-like one-dimensional transistors with SB contacts  has been developed.
The model is free from numerical integration, therefore, it significantly decreases the evaluation times and eases the implementation of the model in Verilog-A.
The developed semi--empirical formulation of the subthreshold current allows to accurately describe the experimental data.
The implementation of the derived analytical $I-V$ expressions in Verilog-A, which is supported by all commercially available circuit simulators, along with a physics-based charge expression will support accurate and predictive simulation needed for designing mixed-signal and analog high-frequency circuits~\cite{Schroter_2013}.
The total current is differentiable throughout all regions of operation.
We have introduced a piece-wise approximation for Fermi--Dirac distribution function and modified the transmission probability using simple elementary functions, which allow to simplify the current calculations. 
Our analytical $I-V$ expressions can be used for the analysis of experimental data as well as for performance predictions for different SB heights, characteristic lengths, and either electron effective mass or band gap of channel material for quasi-1D SB-FETs based on both semiconductor nanowires and nanotubes.

\ifCLASSOPTIONcaptionsoff
  \newpage
\fi



\bibliographystyle{IEEEtran}
\end{document}